\documentclass[conference,compsoc]{IEEEtran}
\IEEEoverridecommandlockouts
\usepackage{cite}
\usepackage{amsmath,amssymb,amsfonts}
\usepackage{algorithmic}
\usepackage[ruled,vlined]{algorithm2e}
\usepackage{graphicx}
\usepackage{textcomp}
\usepackage{xcolor}
\usepackage{enumitem}
\usepackage{kotex}
\usepackage{listings}
\usepackage{subfigure}
\usepackage{xspace}
\usepackage[utf8]{inputenc}
\usepackage{booktabs}
\usepackage{multirow}
\usepackage{float}
\usepackage{makecell}
\usepackage{tikz}
\usepackage{pifont}
\usepackage{ulem}
\usepackage{url}
\usepackage{tabularx}
\usepackage{array}
\usetikzlibrary{arrows.meta, positioning, fit, backgrounds}
\usepackage[hidelinks]{hyperref}
\def\BibTeX{{\rm B\kern-.05em{\sc i\kern-.025em b}\kern-.08em
    T\kern-.1667em\lower.7ex\hbox{E}\kern-.125emX}}
\newcommand{\ours}{\textsc{Anchor}\xspace}

\newif\ifanonymous
\anonymousfalse

\begin{document}
\title{
\ours: Schema-Agnostic Threat Knowledge Graph Construction via Hybrid Ontology Discovery
}
\ifanonymous
\author{%
\IEEEauthorblockN{Anonymous Author(s)}
}
\else
\author{%
\begin{tabular}{c@{\hspace{2em}}c@{\hspace{2em}}c}
Seonwoo Kim & Jinwoo Kim & Daegyu Kang \\
Ministry of National Defense & Incheon International Airport Corporation & Financial Security Institute \\
Seoul, Republic of Korea & Incheon, Republic of Korea & Yongin, Republic of Korea
\end{tabular} \\[2ex]
\begin{tabular}{c@{\hspace{3em}}c}
Daeseong Kim & Insup Lee\textsuperscript{$\ast$} \\
Korean National Police Agency & KAIST \\
Seoul, Republic of Korea & Daejeon, Republic of Korea
\end{tabular}
\thanks{\textsuperscript{$\ast$}Corresponding author: Insup Lee (islee94@kaist.ac.kr)}
}
\fi

\maketitle
\begin{abstract}

Cyber threat intelligence (CTI) reports now serve as essential resources for capturing adversary tactics, techniques, and procedures observed in modern attack campaigns.
While traditional CTI platforms reduce this intelligence to isolated indicators through fixed schemas such as STIX, ontology-based representations preserve the semantic relationships needed for structured threat analysis.
However, existing approaches for ontology-aligned CTI extraction face three challenges: (i) schema-specific pipelines that require manual reconfiguration whenever the schema changes, (ii) prompt-based schema inclusion that fails to scale on large ontologies such as UCO, and (iii) reliance on enterprise LLM APIs that conflicts with privacy constraints when integrating sensitive internal incident data.
In this paper, we present \ours, a \textit{schema-agnostic} CTI knowledge graph construction system that bridges LLMs and formal ontology schemas.
At the core of \ours is \textit{hybrid ontology discovery}, a search-and-navigate mechanism that dynamically explores large-scale ontology schemas, combined with SHACL-based validation to enforce schema-compliant type assignments.
Experimental results on the UCO, STIX, and MALOnt schemas show that \ours outperforms existing baselines in ontology typing and schema compliance.
In addition, \ours with a local LLM closely matches enterprise LLM typing performance, enabling privacy-preserving CTI analysis with high fidelity.

\end{abstract}


\section{Introduction}
\label{sec:introduction}

A comprehensive analysis of cyber campaigns requires more than isolated indicators of compromise (IoC)~\cite{sun2023cyber}.
Cyber threat intelligence (CTI) reports record the threat actors, attack sequences, and technical evidence, explicitly detailing the \textit{causal} relationships between attack steps.
To facilitate automated sharing of such knowledge, 
standardized data formats such as the structured threat information expression (STIX)~\cite{stix21} and the malware information sharing platform (MISP)~\cite{wagner2016misp} have been widely adopted.
However, these indicator-oriented formats primarily capture low-level, isolated data points such as IP addresses and file names, discarding the causal dependencies between attack steps and the analytical reasoning that links them~\cite{gao2021system, li2022attackg, liao2016acing}.

Ontology-based representations address this semantic gap by defining a formal vocabulary of ontology classes, their properties (object and datatype), and structural constraints.
Notable efforts include the unified cyber ontology (UCO)~\cite{syed2016uco}, STUCCO~\cite{iannacone2015developing}, and MALOnt~\cite{rastogi2020malont}.
These frameworks are typically formalized in the web ontology language (OWL) for class hierarchies and properties, supplemented by optional shapes constraint language (SHACL) constraints for structural validation.
However, despite the introduction of many security ontologies over the past decade, no single schema has achieved widespread adoption in practice~\cite{adach2022security}.
Recent studies have therefore explored the automated extraction of ontology-aligned knowledge from unstructured CTI text.
Conventional approaches rely on rule-based pipelines~\cite{husari2017ttpdrill, satvat2021extractor} and classification models~\cite{alam2023looking, peng2025retrieval}.
Unfortunately, these methods depend on hand-crafted rules or fixed type inventories, requiring a complete redesign whenever the target schema changes.
Large language model (LLM)-based methods~\cite{huang2024ctikg, cheng2025ctinexus, huangllm4cti} are better suited for schema-agnostic extraction because they can interpret ontology class descriptions directly without schema-specific engineering.
However, existing LLM-based approaches have been validated only on small custom schemas and exhibit critical limitations when applied to large-scale, real-world ontologies~\cite{buchel2025sok}.

We identify three limitations of existing LLM-based automated ontology extraction approaches.
First, conventional methods~\cite{husari2017ttpdrill, huang2024ctikg, cheng2025ctinexus, huangllm4cti} either define their own custom schemas or hard-code a specific ontology into the extraction pipeline (\textit{schema dependency}).
This rigid design requires a complete redesign whenever the underlying schema changes.
The resolution of schema dependency alone does not solve the scalability problem, as the system must still present the target ontology to the LLM at inference time.
Second, prompt-based schema inclusion strategies fail on large-scale ontologies such as UCO.
Hundreds of hierarchical classes exhaust the context window and degrade the model's ability to distinguish semantically similar types.
Third, an inherent reliance on enterprise LLMs complicates the integration of external CTI with sensitive internal incident data due to privacy and data-sovereignty concerns.
In this context, we derive the following three challenges from the existing literature.

\textbf{Challenge 1) How can we extract ontology-aligned knowledge without being tied to a specific schema?}
To support diverse and evolving security standards~\cite{preuveneers2024ontology}, the system must decouple extraction logic from schema definitions.
This decoupling allows different OWL/SHACL ontologies to be loaded without manually reconfiguring the pipeline.

\textbf{Challenge 2) How can we handle large-scale ontologies that exceed the capacity of prompt-based schema inclusion?}
Rather than including the entire schema in the LLM prompt, the system must dynamically discover and retrieve only task-relevant ontology fragments.
This dynamic retrieval enables schema-aware reasoning within large and hierarchical ontologies while keeping each query within the context window.

\textbf{Challenge 3) How can we support privacy-preserving CTI analysis with local LLMs?}
External CTI provides greater operational value when integrated with internal alerts, logs, and incident reports, but such data is often too sensitive to expose to enterprise LLMs.
The system must therefore enable ontology-grounded reasoning with locally deployed open-source LLMs while preserving competitive typing performance.


To this end, we propose \ours (\underline{A}daptive \underline{N}avigation for \underline{C}ybersecurity \underline{H}ybrid \underline{O}ntology \underline{R}easoning), a \textit{schema-agnostic} system that builds a structured threat knowledge graph from unstructured CTI reports.
The proposed method presents a \textit{hybrid ontology discovery} mechanism, which enables LLMs to dynamically explore relevant sub-graphs of complex ontologies and separates the extraction pipeline from any specific schema.
To ensure accurate and schema-compliant ontology alignment, \ours also considers a SHACL-based validation during ontology typing.
The main contributions of this paper are summarized as follows.

\begin{itemize}
\item We propose \ours, a schema-agnostic CTI knowledge graph construction framework that decouples the extraction pipeline from any specific ontology schema, supporting arbitrary OWL/SHACL ontologies at runtime without manual reconfiguration.

\item We introduce \textit{hybrid ontology discovery}, which combines embedding-based semantic search with LLM-guided recursive navigation to retrieve only task-relevant ontologies, integrated with closed-loop SHACL validation to ensure schema-compliant knowledge graphs.

\item We demonstrate privacy-preserving CTI knowledge graph construction with locally deployed open-source LLMs, retaining 99.2\% (entity) and 97.8\% (predicate) of the best enterprise LLM's performance.
\end{itemize}

To foster research on the topic, we make \ours publicly available (https://github.com/seonwookim92/acsac-ANCHOR).


\section{Background}\label{sec:background}

This section reviews the necessary background on CTI data formats, ontologies, and LLM-based knowledge extraction.

\subsection{CTI Data Formats and Ontologies}
\label{sec:background_cti}

\textcolor{black}{CTI records adversary tactics, techniques, and procedures (TTPs) and IoCs.
This intelligence serves as a shared resource for organizations to coordinate defenses against rapidly changing threats~\cite{wagner2019cyber}.
Open-source CTI (OSCTI) is collected from security blogs, threat reports, and vulnerability databases, providing a critical resource for understanding dynamic threat environments.
To standardize the exchange of threat information, the security community widely adopts three data formats and protocols.
STIX~\cite{stix21} defines a set of domain objects (e.g., threat actors, malware, vulnerabilities) and their relationships in a JSON-based format.
MISP~\cite{wagner2016misp} provides a collaborative framework to share IoCs using a predefined attribute taxonomy.
The trusted automated exchange of intelligence information (TAXII)~\cite{taxii} acts as the transport protocol to distribute STIX-formatted data between organizations.}

\begin{figure}[t]
    \centering
    \includegraphics[width=0.9\columnwidth]{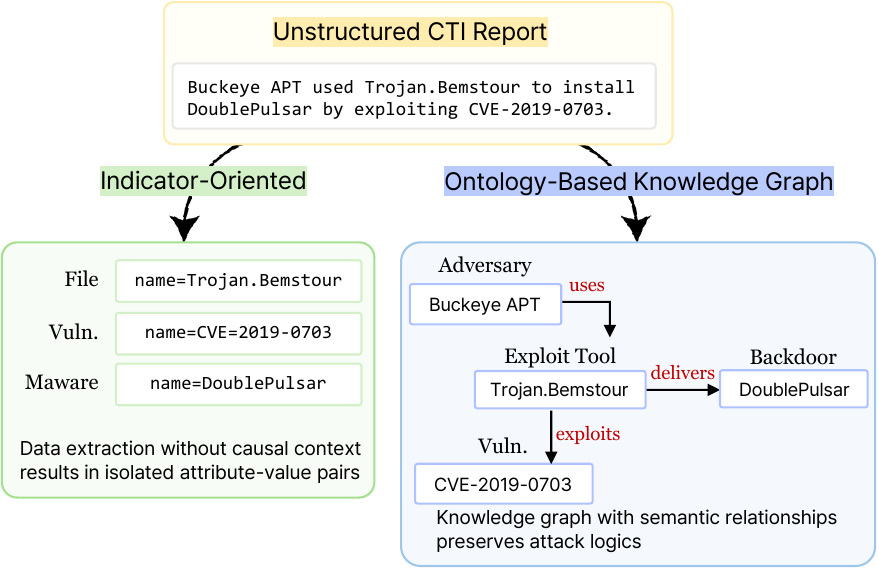}
    \caption{
           Comparison of two CTI data formats: indicator-oriented flat data and ontology-based knowledge graph.
           }
    \label{fig:flat_vs_ontology}
\end{figure}

\textcolor{black}{While these formats enable efficient indicator sharing, their structures are limited to flat attribute-value pairs and predefined relationship types~\cite{gao2021system}.
Fig.~\ref{fig:flat_vs_ontology} illustrates differences between this indicator-oriented view and an ontology-based knowledge graph.
In the threat knowledge graph, semantic relationships connect security entities into a coherent structure.
This connectivity allows analysts to trace the full attack flow rather than examine individual indicators in isolation.
Ontologies are typically expressed in the web ontology language (OWL), which defines class hierarchies alongside object and datatype properties.
These definitions often include optional shapes constraint language (SHACL) constraints for structural validation, such as required attributes and cardinality bounds.}

\textcolor{black}{UCO~\cite{syed2016uco} unifies concepts from multiple security standards (e.g., CyBOK, NIST, MITRE ATT\&CK) into a single hierarchical framework defined in OWL/SHACL, containing hundreds of classes and properties.
In contrast, MALOnt~\cite{rastogi2020malont} provides a lightweight malware-focused ontology with only 75 classes, 10 relations, and 12 properties.
Despite the studies on security ontologies, the construction and maintenance of a comprehensive cybersecurity ontology remains an open problem.
A recent survey reports inconsistencies and ambiguities in cybersecurity terminology that hinder communication between security professionals~\cite{adach2022security}.
Furthermore, no existing ontology fully aligns with the major security standards required for seamless cross-platform integration.}
These practical limitations motivate the scenarios described in Section~\ref{sec3}.

\subsection{LLM-based Knowledge Extraction}
\label{sec:background_llm_mcp}

\textcolor{black}{LLMs have demonstrated strong performance in natural language understanding and information extraction tasks~\cite{brown2020language, achiam2023gpt}.
In the CTI domain, LLMs offer a practical alternative to traditional rule-based or keyword-matching systems.
While these traditional systems struggle to capture the intent and contextual nuances of attacker behavior, recent studies~\cite{li2022attackg, huang2024ctikg, cheng2025ctinexus, huangllm4cti} show that LLMs can extract key entities and their semantic relationships from threat reports with high performance.
This automated extraction reduces the need for manual analyst intervention in knowledge graph construction.}

The integration of LLMs with external knowledge sources, such as databases and ontology schemas, has traditionally required separate implementations for each data source.
This requirement creates tightly coupled pipelines that are difficult to maintain.
Recent agent frameworks address this fragmentation by providing standardized interfaces to access external knowledge dynamically.
This interactive approach grounds LLM outputs in verifiable references rather than relying solely on parametric knowledge.
Our design adopts this methodology to ensure flexibility and maintainability.
A uniform interface between the LLM agent and the external ontology, aligned with open standards such as the model context protocol (MCP)~\cite{hou2025model}, successfully decouples the extraction logic from any specific schema (Section~\ref{sec4}).

\section{Motivating Examples}
\label{sec3}

This section illustrates the practical limitations of current automated CTI extraction through two scenarios: 
(i) information loss issues common to all indicator-oriented pipelines and (ii) a scalability limitation caused by prompt-based schema inclusion in LLM.

\subsection{Information Loss in Indicator-oriented Formats}
\label{sec31}

\textcolor{black}{
To illustrate the information loss caused by indicator-oriented CTI data, consider the following passage from typical threat analysis reports:}
\begin{quote}
\textit{``In early March 2024, security analysts identified a spear-phishing campaign targeting Southeast Asian financial institutions. The attacker delivered a Microsoft Word document disguised as an invoice via email. When a user opens the document, a macro executes to download a secondary payload from hxxp://update\allowbreak{}-check[.]site/\allowbreak{}loader.exe. This payload maintained access by creating a scheduled task named ``WindowsUpdateCheck'' and afterward communicated with a command and control (C2) server at 45.77.23.91 via TCP port 443. Analysis revealed that these techniques align with the activities of APT-X, known for macro-based initial access and invoice-themed lures.''}
\end{quote}

When this scenario is mapped into indicator-oriented formats such as STIX, most CTI platforms extract only a limited subset of isolated indicators:
\begin{itemize}
    \item \textbf{Indicators:}
    \begin{itemize}
        \item URL: ``update-check[.]site''
        \item IP: ``45[.]77[.]23[.]91''
        \item File: ``loader.exe''
    \end{itemize}
    \item \textbf{Objects:}
    \begin{itemize}
        \item Scheduled Task: ``WindowsUpdateCheck''
        \item Identity: \textit{Financial Sector}
    \end{itemize}
    \item \textbf{Relationships:} 
    \begin{itemize}
        \item Attributed-to \textit{APT-X}
    \end{itemize}
    
\end{itemize}
\textcolor{black}{
Note that this isolation leads to the loss of semantic connectivity: indicator-oriented formats log discrete values while omitting the causality and rationale that link them together.
For instance, the passage above includes a causal chain (e.g., (i) macro execution downloads the secondary payload and (ii) the C2 connection enables remote control) and an attribution rationale (TTP overlap with APT-X), yet the STIX output preserves only flat identifiers without any of this contextual reasoning.
As a result, the structured output captures a fragmented snapshot of the original report, with no representation of the attack logic that connects these indicators.}
To preserve these crucial relationships, an ontology-based knowledge graph offers a promising alternative that explicitly models semantic connectivity between threat entities.

\subsection{Limited Scalability of Prompt-Based Schema Inclusion}
\label{sec32}

\textcolor{black}{
Recent LLM-based methods map entities to ontologies using prompt-based schema inclusion, which embeds the entire class list directly into the input prompt.}
\textcolor{black}{
As shown in Table~\ref{tab:motivation-comparison}, existing systems~\cite{husari2017ttpdrill, huang2024ctikg, cheng2025ctinexus, huangllm4cti} either hard-code a specific schema, operate without one, or rely on prompt-based inclusion at a small scale (60 to 77 elements). While prompt-based inclusion is feasible for small schemas like MALOnt and STIX, it introduces three severe limitations when applied to large-scale ontologies such as UCO (419 classes):
\begin{itemize}[leftmargin=*, topsep=2pt, itemsep=1pt]
    \item \textbf{Context exhaustion:} Massive prompts with hundreds of class descriptions trigger the ``Lost in the Middle'' phenomenon~\cite{liu2024lost}, causing the model to miss relevant context.
    \item \textbf{Precision degradation:} An excessive number of candidates degrades semantic precision, causing the model to confuse adjacent types (e.g., \textit{Process}, \textit{Action}, \textit{Event}) or hallucinate non-existent classes.
    \item \textbf{Maintenance burden:} A fixed class list requires manual reconfiguration whenever the schema is updated.
\end{itemize}
Furthermore, none of the existing baselines verify whether assigned types satisfy formal structural constraints, allowing invalid mappings to enter the knowledge graph undetected.}
\textcolor{black}{
These scalability and validation gaps highlight a critical misalignment between static LLM prompts and complex security ontologies.
To process large-scale schemas without context exhaustion, a system requires dynamic exploration that retrieves only task-relevant fragments on demand.
Furthermore, to prevent hallucinated mappings from corrupting the output, the system must enforce formal constraints before committing the data.
Motivated by these exact requirements, Section~\ref{sec4} introduces the architectural design of \ours, which replaces prompt-based full schema inclusion with an effective ontology discovery and validation mechanism.
}

\begin{table}[t]
\centering
\caption{Comparison of CTI Knowledge Graph Construction Systems}
\label{tab:motivation-comparison}
\resizebox{\columnwidth}{!}{
\begin{tabular}{lccc}
\toprule
\textbf{System} & \textbf{Method} & \textbf{Schema dependency} & \textbf{Schema size} \\
\midrule
TTPDrill~\cite{husari2017ttpdrill} & Rule       & Fixed (ATT\&CK)        & - \\
CTIKG~\cite{huang2024ctikg}        & LLM (prompt)  & Open-ended             & - \\
CTINexus~\cite{cheng2025ctinexus}  & LLM (prompt)  & Fixed (MALOnt) & 60 \\
LLM4CTI~\cite{huangllm4cti}        & LLM (prompt)  & Fixed (STIX)         & 77 \\
\midrule
\textbf{\ours}                     & \textbf{LLM (hybrid)} & \textbf{Any OWL/SHACL} & \textbf{Up to 997 (UCO)} \\
\bottomrule
\end{tabular}}
\end{table}

\begin{figure*}[t]
  \centering
  \includegraphics[width=0.95\textwidth]{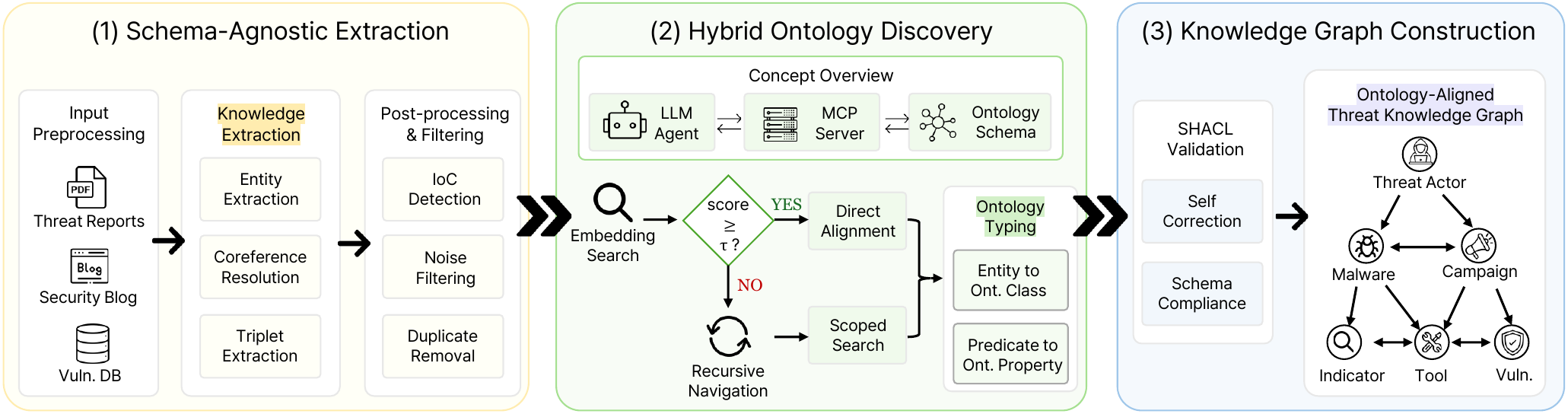}
\caption{      
     Architecture of the \ours System: 
    (i) Schema-agnostic extraction preprocesses CTI documents and extracts entities, coreferences, and triplets without binding to a fixed ontology schema; 
    (ii) Hybrid ontology discovery maps each entity and predicate to an ontology class or property URI via embedding search when confidence meets $\tau$, or recursive navigation otherwise;
    (iii) Knowledge graph construction applies SHACL validation with closed-loop self-correction to build an ontology-aligned threat knowledge graph.
} 
  \label{fig:architecture}
\end{figure*}

\section{\ours Design}
\label{sec4}

We propose \ours, a schema-agnostic threat knowledge graph construction system that aligns entities and predicates to large-scale OWL/SHACL ontologies via hybrid ontology discovery.
This section describes the system overview, the schema-agnostic extraction pipeline, and the hybrid ontology discovery procedure.

\subsection{System Overview}
\label{sec41}

\ours aims to construct threat knowledge graphs from fragmented CTI data based on the following three principles:
(i) the extraction pipeline is decoupled from any specific ontology schema, supporting arbitrary OWL/SHACL ontologies at runtime;
(ii) ontology fragments are discovered \textit{dynamically} rather than included as a whole in the LLM prompt, keeping each query within the context window;
(iii) during threat knowledge graph construction, every type assignment is verified against formal schema constraints before commitment, ensuring schema compliance.
As shown in Fig.~\ref{fig:architecture}, the resulting architecture consists of three core components: schema-agnostic extraction, hybrid ontology discovery, and knowledge graph construction.

\textbf{Schema-agnostic extraction.}
This stage extracts entities, coreferences, and relation triplets from preprocessed CTI text without binding the extraction logic to any specific ontology, and forwards the structured output to the next stage for type assignment.
Section~\ref{sec42} details the full extraction pipeline.

\textbf{Hybrid ontology discovery.}
This stage aligns each extracted element to a formal ontology class or property URI by combining embedding-based semantic search with LLM-guided hierarchical navigation over the target OWL/SHACL schema.
When the search confidence falls below a predefined threshold, the system switches to recursive schema traversal, ensuring that domain-specific or novel terminology is correctly resolved.
Section~\ref{sec43} details the discovery algorithm.

\textbf{Knowledge graph construction.}
The validated type assignments are assembled into an ontology-aligned knowledge graph through a SHACL-based closed-loop correction process: each candidate assignment is checked against the target schema, and the mapper is re-invoked with the violation report until the output is conformant or a retry budget is exhausted. The resulting graph is serialized as JSON and, optionally, as a Neo4j-importable Cypher file to support downstream multi-hop queries over attack chains.

\subsection{Schema-Agnostic Extraction}
\label{sec42}


To prepare unstructured CTI data for ontology-aligned knowledge extraction, \ours processes raw inputs through a sequential pipeline: (i) input preprocessing, (ii) knowledge extraction, and (iii) post-processing and filtering.

\subsubsection{Input Preprocessing}
\label{sec421}

Given unstructured CTI text such as threat reports, \ours  applies a recursive character-based splitting strategy to handle documents that exceed the LLM context window.
The splitter respects natural semantic boundaries (paragraph breaks, sentence boundaries, word boundaries) and retains a fixed-size overlap window between consecutive chunks to preserve cross-boundary context.

\subsubsection{Knowledge Extraction}
\label{sec422}

After preprocessing, \ours extracts structured entities and relationships from preprocessed chunks through three phases: entity extraction, coreference resolution, and triplet extraction.

\textbf{Entity extraction.}
\ours identifies all named entities from each chunk through a dedicated LLM extraction call.
An entity here refers to any phrase that denotes a concrete object or referent in the text, in contrast to descriptive expressions (e.g., adjectives, adverbs) or behavioral phrases that do not stand as standalone referents.

Rather than restricting the process to predefined entity classes, the system captures a broad spectrum of referents and defers formal ontology type assignment to the subsequent discovery stage.
Each extracted entity is represented as a tuple $(\textit{name},\, \textit{type\_hint},\, \textit{properties})$,
where \textit{type\_hint} is a short phrase that guides ontology class search, and \textit{properties} captures
clearly stated literal attributes (e.g., implementation language, alias, first-seen date).
To maintain consistent referencing, the extraction enforces name normalization (e.g., stripping leading articles, lowercasing) and performs cross-chunk unification using a normalized key, so that surface variations such as ``Fancy Bear Hacking Group'' and ``Fancy Bear'' map to the same canonical entry while non-entity behavioral descriptions are filtered out.

\textbf{Coreference resolution.}
After entity extraction, \ours applies an entity-aware coreference resolution pass to each chunk.
The LLM replaces anaphoric references (pronouns, role descriptors, near-demonstratives) with the corresponding canonical entity name from the unified list, reducing misattributed or missing relations caused by unresolved anaphora.

\textbf{Triplet extraction.}
With the extracted entity list in place, the system extracts structured relationships per chunk, providing the entity list as an explicit restriction so that only known, verified entities appear as relation endpoints.
This two-pass design (entities first, relations second) eliminates subject-bias (over-generating relations for prominent entities while neglecting less salient ones) and dangling references (relation endpoints absent from the final inventory), both of which frequently arise when entity discovery and relation extraction are performed simultaneously.
ObjectProperty triplets represent entity-to-entity relations of the form $(e_s, p, e_o)$ with $p$ a concise predicate phrase (e.g., ``uses'', ``targets'', ``exploits'', ``attributed to'') and an evidence sentence from the source text attached for grounding context during ontology alignment; DatatypeProperty triplets capture entity-to-literal associations such as language strings, timestamps, or boolean flags.
The LLM is instructed to be exhaustive, evaluating every entity pair and capturing all stated attributes regardless of salience.

\subsubsection{Post-processing and Filtering}
\label{sec423}

Following knowledge extraction, \ours applies three deterministic post-processing steps: IoC detection, noise filtering, and duplicate removal. 
In IoC detection, \ours matches entity names against regular expression patterns and overrides their type hints with a schema-agnostic descriptor (e.g., ``IPv4 address'', ``SHA-256 hash''), enabling reliable class resolution in the mapping stage regardless of which ontology is active. 
In noise filtering, \ours removes non-entity strings such as temporal expressions, short strings, and common descriptors. 
Finally, in duplicate removal, \ours merges duplicate entities sharing the same normalized key and removes duplicate or dangling triplets. The filtered output is then forwarded to the subsequent hybrid ontology discovery stage.

\subsection{Hybrid Ontology Discovery}
\label{sec43}

With the schema-agnostic knowledge extraction complete, \ours aligns each extracted element to a formal ontology class and property.
Hybrid ontology discovery provides this alignment through a uniform tool-based interface between the LLM agent and any OWL/SHACL ontology.
At initialization, \ours parses the target ontology file, pre-computes embeddings for all class names and descriptions, and indexes the type hierarchy.
It exposes six tools covering three targets (classes, attributes, relations), each with two operations: (i) embedding-based search that retrieves candidates by semantic similarity, and (ii) hierarchical recursive navigation that the LLM invokes when search confidence is below the threshold $\tau$.
\footnote{To facilitate broader reuse, the discovery suite is provided as an MCP-compatible server that includes the six core tools used in this work along with auxiliary functions for general-purpose ontology exploration. The underlying hybrid ontology discovery architecture, however, is agnostic to any specific tool-call protocol.}

\subsubsection{Entity Ontology Typing}
\label{sec431}

\ours determines the appropriate formal ontology class for each extracted entity, representing it as a uniform resource identifier (URI), by executing Algorithm~\ref{alg:hybrid}. 
For this phase, the algorithm is instantiated using class-specific semantic search, the root class set $U_{\text{roots}}$, and owl:Thing as the default fallback. 

Unlike predicate ontology typing, entity ontology typing executes only Steps 1 and 2 of the discovery process, with Algorithm~\ref{alg:hybrid} instantiated in \textsc{Entity} mode.

\begin{algorithm}[t]

\caption{Hybrid Ontology Discovery}
\label{alg:hybrid}
\begin{algorithmic}[1]
 \renewcommand{\algorithmicrequire}{\textbf{Input:}}
  \renewcommand{\algorithmicensure}{\textbf{Output:}}
\REQUIRE Query hint $h$; root URIs $U_{\text{roots}}$; confidence threshold $\tau$; discovery mode $m \in \{\textsc{Entity}, \textsc{Predicate}\}$.
\ENSURE Matched ontology element URI $u^{*}$.

\STATE \textit{/* Step 1: Embedding-based search */}
\STATE $U_{\text{cands}} \gets$ Search($h$)
\IF{$U_{\text{cands}} \neq \emptyset$ \AND score(top($U_{\text{cands}}$)) $\geq \tau$}
  \RETURN top($U_{\text{cands}}$)
\ENDIF

\STATE \textit{/* Step 2: Hierarchical recursive navigation */}
\STATE $u_{\text{curr}} \gets$ LlmSelect($h$, $U_{\text{roots}}$, RetrieveDesc($U_{\text{roots}}$))
\IF{$u_{\text{curr}} = \text{none}$}
    \RETURN Fallback($m$)
\ENDIF
\STATE $u_{\text{scope}} \gets \emptyset$
\WHILE{\textbf{true}}
\STATE $S \gets$ Children($u_{\text{curr}}$)
\IF{$S = \emptyset$}
    \STATE $u_{\text{scope}} \gets u_{\text{curr}}$; \textbf{break}
\ENDIF
\STATE $u_{\text{best}} \gets$ LlmSelect($h$, $S$, RetrieveDesc($S$))
\IF{$u_{\text{best}} = \text{none}$}
    \STATE $u_{\text{scope}} \gets u_{\text{curr}}$; \textbf{break}
\ENDIF
\STATE $u_{\text{curr}} \gets u_{\text{best}}$
\ENDWHILE

\IF{$m = \textsc{Entity}$}
    \RETURN $u_{\text{scope}}$
\ENDIF

\STATE \textit{/* Step 3: Scoped embedding search (predicate) */}
\IF{$u_{\text{scope}} \neq \emptyset$}
    \STATE $U_{\text{cands}} \gets$ Search($h$, scope = $u_{\text{scope}}$)
    \IF{$U_{\text{cands}} \neq \emptyset$ \AND score(top($U_{\text{cands}}$)) $\geq \tau$}
        \RETURN top($U_{\text{cands}}$)
    \ENDIF
\ENDIF

\RETURN Fallback($m$)
\end{algorithmic}
\end{algorithm}

\textbf{Step~1: Embedding-based search.}
The system first performs embedding-based search, which computes a relevance score for each candidate URI $u$ in the ontology graph $G$:

\begin{equation}
\mathrm{score}(u) = \frac{1}{2}\,\mathrm{sim}(\mathbf{q}, \mathbf{e}^{name}_{u})
                  + \frac{1}{2}\,\mathrm{sim}(\mathbf{q}, \mathbf{e}^{desc}_{u})
                  + \delta_k\,,
\label{eq:class_score}
\end{equation}

where $\mathbf{q}$ is the query embedding derived from the query hint $h$ (e.g., the entity's \textit{type\_hint}), $\mathbf{e}^{name}_{u}$ and $\mathbf{e}^{desc}_{u}$ are pre-computed embeddings of the class name and description, $\mathrm{sim}(\cdot)$ denotes cosine similarity, and $\delta_k$ is a fixed keyword bonus (set to $0.3$ in our experiments) applied when the query string appears verbatim in the class name or description.
If the top candidate's score meets or exceeds the entity confidence threshold $\tau_{\text{entity}}$ (set to $0.45$ in our experiments), the class is selected immediately.

\textbf{Step~2: Hierarchical recursive navigation.}
If no candidate from Step~1 clears the threshold $\tau_{\text{entity}}$, the system activates hierarchical recursive navigation.
From the set of root classes $U_{\text{roots}}$, the LLM evaluates semantic definitions of each subclass tier and selects the most logically matching branch, drilling down iteratively until it reaches a leaf node or determines that no sufficiently matching branch remains.
This top-down approach compensates for the limitations of embedding models by leveraging LLM reasoning, enabling precise classification of domain-specific or novel terminology.
The LLM bases this decision on the candidate subclasses' names, textual descriptions, and child counts presented at each tier, signaling termination when no candidate aligns with the target query hint $h$.

\textbf{Fallback.} If the recursive navigation fails to identify a suitable class, the entity defaults to owl:Thing, corresponding to the fallback return in Algorithm~\ref{alg:hybrid}. This defensive assignment ensures the extracted entity and its associated relations remain in the knowledge graph without introducing an incorrect or unverified type.

\subsubsection{Predicate Ontology Typing}
\label{sec432}

To assign a formal property URI to each extracted triplet, \ours extends the search-and-navigate strategy used in entity ontology typing. 
Depending on the triplet type, it determines either an ObjectProperty URI for entity-to-entity relations or a DatatypeProperty URI for entity-to-literal attributes. 

Unlike entity ontology typing, this phase utilizes all three steps of Algorithm~\ref{alg:hybrid}, 
with the algorithm instantiated in \textsc{Predicate} mode.

\textbf{Step~1: Embedding-based search.}
\ours deploys two distinct search tools to handle attributes and relations. 
For literal attributes, the embedding-based search identifies the optimal DatatypeProperty. It extends Equation~\ref{eq:class_score} with a datatype inference heuristic, applying a score boost when the inferred XSD type (e.g., xsd:dateTime, xsd:integer) aligns with the property's declared XML Schema Definition (XSD) range. 
For entity connections, the relation search targets ObjectProperty URIs by utilizing the evidence sentence captured during triplet extraction as additional context. It applies direction-aware score adjustments, adding a bonus ($0.10$) for direct forward relations and a minor penalty ($-0.05$) for inverse relations. 
To ensure completeness, both tools evaluate properties inherited from the transitive superclass hierarchy, leveraging SHACL domain annotations back-propagated during schema loading.

\textbf{Step~2: Hierarchical recursive navigation.}
If the search score falls below the predicate confidence threshold $\tau_{\text{predicate}}$ (set to $0.30$), the system triggers LLM-guided traversal. 
The traversal starting point depends on the target: data-property navigation roots at the bound entity URI, while object-property navigation roots at the $(s, o)$ URI pair. 
From these roots, the LLM navigates a structured property list organized by inheritance level and domain class. For object properties, the LLM additionally infers the correct assertion direction (forward or inverse).

\textbf{Step~3: Scoped embedding search.}
Large schemas often group properties under intermediate classes, making exhaustive LLM traversal impractical (e.g., UCO contains 578 properties). 
When navigation reveals a collapsed property group anchored to an intermediate class $u_{\text{scope}}$, \ours re-executes the embedding similarity search restricted solely to the properties of that class. 
This scoped pass allows the system to handle large hierarchies efficiently once anchored to a specific branch.

\textbf{Fallback.}
If all three steps fail to identify a matching property, \ours defaults to rdfs:label for literal attributes and rdfs:seeAlso for entity relations. 
Similar to entity ontology typing, this defensive mapping ensures every extracted triplet is preserved in the output graph without fabricating incorrect schema assignments.

\subsection{Knowledge Graph Construction with SHACL Validation}
\label{sec44}

After finalizing the entity and predicate mappings, \ours assembles the typed elements into a unified knowledge graph. 
To ensure structural integrity, the system applies a SHACL-based validation mechanism that evaluates every entity against the node shapes defined in the target ontology. 
This validation targets cardinality constraints, verifying that required properties are present and that value counts fall within declared bounds. 
Because LLM-based extraction often omits mandatory attributes, missing required properties represent the most common source of schema violations in this domain.

For example, under the UCO schema, a cardinality violation occurs if an extracted malware entity lacks the required hash algorithm attribute, a common omission that renders the indicator unusable for downstream matching. 
Upon detecting a violation, the system generates a structured feedback message detailing the violating entity, the failed constraint, and the expected correction. 
The LLM agent receives this feedback and re-invokes the appropriate discovery tool to rectify the omission (e.g., searching for a missing required property). 
\ours limits this closed-loop self-correction to three retries to prevent unbounded execution. 
If the violation persists after the maximum attempts, the system retains the entity with its assigned class and flags it with a validation warning rather than silently discarding it. 
This approach minimizes unverified type assignments while preserving the extracted intelligence.

Beyond constraint checking, these SHACL shapes facilitate facet discovery for ontologies that organize auxiliary properties into compound structures (e.g., UCO). 
The system performs this through a multi-strategy lookup combining property domain inspection, inheritance traversal, and SHACL shape resolution. 
Finally, \ours serializes the validated type assignments as an ontology-aligned knowledge graph in JSON format.

\begin{figure}[t]
    \centering
    \includegraphics[width=0.95\columnwidth]{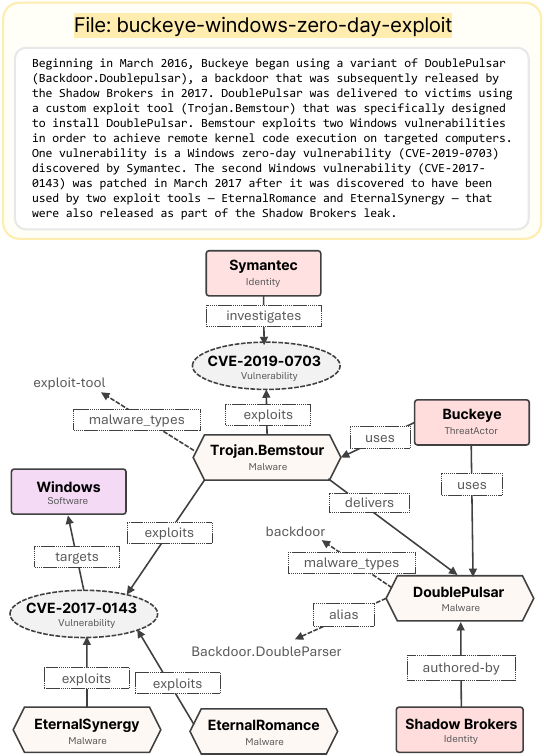}
    \caption{
    Example knowledge graph constructed by \ours from a Buckeye APT campaign report.
    }
    \label{fig:kg_example}
\end{figure}

To provide an intuitive understanding of how \ours reconstructs semantic connectivity from an unstructured CTI report, as shown in Fig.~\ref{fig:kg_example}, we visualize an example knowledge graph constructed from a Buckeye APT campaign report.
The resulting graph contains 10 typed entities of five classes (ThreatActor, Identity, Malware, Software, and Vulnerability), connected by typed predicates (e.g., uses, delivers, and exploits) that trace the attack chain from the exploit tool through the vulnerabilities and the backdoor to the Shadow Brokers leak.
The two vulnerabilities (CVE-2019-0703 and CVE-2017-0143) are typed against the Vulnerability class, demonstrating accurate resolution of numerical CVE references.
Datatype properties such as malware\_types (``backdoor'' and ``exploit-tool'') and alias (``Backdoor.DoubleParser'') attach literal attributes to the entities.
These typed entities, properties, and predicates jointly reconstruct the semantic connectivity discussed in Section~\ref{sec31}, including the causal chain, the temporal sequence, and the qualitative reasoning that indicator-oriented formats omit.


\section{Evaluation}
\label{sec:evaluation}

For reproducibility, we use a benchmark provided by CTINexus~\cite{cheng2025ctinexus}, consisting of 149 CTI reports with manually annotated entities, triplets, and ontology type labels. 
\textcolor{black}{The original benchmark targets only the small-scale MALOnt schema, so we additionally reconstruct ground-truth type labels for two larger schemas, UCO and STIX 2.1, as summarized in Table~\ref{tab:schema_stats}.}
\textcolor{black}{Three cybersecurity researchers established the ground truth through a human-in-the-loop process.}
\textcolor{black}{An ensemble of three LLMs (GPT-5.4, Claude-Sonnet-4-6, and Gemini-3.1-flash) produced the initial candidates, and the researchers manually cross-examined and resolved conflicting assignments by consensus.}

\begin{table}[t]
\centering
\caption{Statistics of Target Ontology Schemas}
\label{tab:schema_stats}
\begin{tabular}{lccc}
\toprule
\textbf{Schema} & \textbf{Classes} & \textbf{Relations} & \textbf{Properties} \\
\midrule
UCO & 419 & 177 & 578 \\
STIX 2.1 & 109 & 92 & 311 \\
MALOnt & 75 & 10 & 12 \\
\bottomrule
\end{tabular}
\end{table}

\textcolor{black}{For our baselines, we compare the F1 scores of \ours against three systems: TTPDrill~\cite{husari2017ttpdrill}, CTINexus~\cite{cheng2025ctinexus}, and LLM4CTI~\cite{huangllm4cti}.}
\textcolor{black}{We exclude CTIKG~\cite{huang2024ctikg}, as it performs only knowledge extraction without ontology typing.}
\textcolor{black}{Since CTINexus originally types only entities, we extend it to predicate typing by reusing its prompt-based entity ontology typing method on relation predicates.}
\textcolor{black}{We deploy \ours on a workstation equipped with an NVIDIA GB10 board and 128~GB of memory, where we serve Qwen3.5-35B locally via vLLM~\cite{kwon2023efficient}.}
\textcolor{black}{For a fair comparison, all baselines also use Qwen3.5-35B as their underlying model.}
\textcolor{black}{In the comparison against enterprise LLMs (Section~\ref{sec55}), we additionally use GPT-5.4-mini and Claude Haiku-4.5.}

\subsection{Ontology Typing Performance}
\label{sec52}

We examine the ontology typing performance on three schemas with different scales: UCO (large, 419 classes), STIX (medium, 109 classes), and MALOnt (small, 75 classes).
Ontology typing measures whether the extracted entities and relations can be aligned to formal ontology classes and properties.
\textcolor{black}{Since the UCO schema is deeply nested with multi-level class hierarchies, we adopt a hierarchical F1 score that assigns full credit (1.0) for an exact match and partial credit (0.6 for a one-step parent or child mismatch, 0.3 for a two-step mismatch) to capture semantic proximity.}
\textcolor{black}{We evaluate two complementary tasks: entity ontology typing, where each extracted entity is mapped to an ontology \textit{class} uniform resource identifier (URI), and predicate ontology typing, where each extracted relation predicate is mapped to an ontology \textit{property} URI.}

\textbf{Entity ontology typing.}
\textcolor{black}{For each extracted entity, the system selects the best-matching class URI from the target ontology.
As shown in Table~\ref{tab:entity_typing}, \ours achieves the highest performance on every schema, with an average F1 of 0.7371.
Specifically, \ours outperforms the second-best baseline (LLM4CTI) by 62.5\% (from 0.4521 to 0.7347) on the UCO schema.
The margin shrinks on smaller schemas: \ours outperforms LLM4CTI by 10.6\% (from 0.7886 to 0.8724) on STIX and by 0.7\% (from 0.6891 to 0.6942) on MALOnt.
CTINexus, which includes the entire schema in the LLM prompt, suffers a 30.3\% performance drop when scaling from MALOnt (0.6370) to UCO (0.4439).
TTPDrill never exceeds 0.2305 on any schema because it relies on a simple rule-based pipeline.}

\begin{table}[t]
    \centering
    \renewcommand\arraystretch{1.0}
    \caption{Entity Ontology Typing Performance on Three Schemas}
    \begin{center}
    \begin{tabular}{lcccc}
    \toprule
    \textbf{System} & \textbf{UCO} & \textbf{STIX} & \textbf{MALOnt} & \textbf{Average} \\
    \midrule
    TTPDrill~\cite{husari2017ttpdrill}  & 0.0652 & 0.1042 & 0.2305 & 0.1333 \\
    CTINexus~\cite{cheng2025ctinexus}   & 0.4439 & 0.5698 & 0.6370 & 0.5502 \\
    LLM4CTI~\cite{huangllm4cti}         & 0.4521 & 0.7886 & 0.6891 & 0.6432 \\
    \ours                               & \textbf{0.7347} & \textbf{0.8724} & \textbf{0.6942} & \textbf{0.7371} \\
    \bottomrule
    \end{tabular}
    \label{tab:entity_typing}
    \end{center}
\end{table}

\textcolor{black}{We observe that this performance gap widens as the schema size grows, highlighting the contrast between prompt-based schema inclusion and the dynamic hybrid ontology discovery provided by \ours.
As the number of candidate classes increases from 75 in MALOnt to 419 in UCO, prompt-based baselines exhaust the LLM context window and degrade due to the ``lost in the middle'' phenomenon~\cite{liu2024lost}. 
In contrast, \ours retrieves only the relevant ontology fragments and remains stable on the larger schema.
The small performance difference on MALOnt indicates that hybrid ontology discovery offers limited benefit when the schema fits easily within a single prompt. 
However, its advantage becomes highly pronounced on large schemas such as UCO, where prompt-based inclusion fails.}

\begin{table}[t]
    \centering
    \renewcommand\arraystretch{1.0}
    \caption{Predicate Ontology Typing Performance on Three Schemas}
    \begin{center}
    \begin{tabular}{lcccc}
    \toprule
    \textbf{System} & \textbf{UCO} & \textbf{STIX} & \textbf{MALOnt} & \textbf{Average} \\
    \midrule
    TTPDrill~\cite{husari2017ttpdrill}  & 0.0033 & 0.0000 & 0.0000 & 0.0011 \\
    CTINexus~\cite{cheng2025ctinexus}   & 0.1282 & 0.4289 & 0.4528 & 0.3366 \\
    LLM4CTI~\cite{huangllm4cti}         & 0.4000 & 0.5355 & \textbf{0.5647} & 0.5001 \\
    \ours                               & \textbf{0.5180} & \textbf{0.5860} & 0.5412 & \textbf{0.5484} \\
    \bottomrule
    \end{tabular}
    \label{tab:predicate_typing}
    \end{center}
\end{table}

\textbf{Predicate ontology typing.}
The system maps each extracted relation predicate to a formal property URI (either ObjectProperty or DatatypeProperty).
\textcolor{black}{As shown in Table~\ref{tab:predicate_typing}, \ours achieves the highest average F1 of 0.5484, outperforming the baselines on the UCO and STIX schemas.}
\textcolor{black}{Specifically, \ours outperforms LLM4CTI by 29.5\% (from 0.4000 to 0.5180) on UCO and by 9.4\% (from 0.5355 to 0.5860) on STIX.}
On MALOnt, \ours scores 0.5412, which is 4.3\% below LLM4CTI at 0.5647.
TTPDrill records zero performance on STIX and MALOnt, since its template-based approach cannot produce property URIs outside the predefined ATT\&CK vocabulary.

\textcolor{black}{Predicate ontology typing is more difficult than entity ontology typing for all four systems: \ours drops by 25.6\% (from 0.7371 to 0.5484) and LLM4CTI drops by 22.2\% (from 0.6432 to 0.5001).}
\textcolor{black}{The gap is intuitive since a predicate's correct property URI depends not only on the surface verb but also on the domain and range of its subject and object entities, and on the direction of the edge (e.g., \textit{uses} vs.\ \textit{used\_by}).}
\textcolor{black}{Single-word embedding similarity alone is therefore insufficient, and the schema-navigation component of hybrid ontology discovery becomes the main contributor to predicate ontology typing, as analyzed in Section~\ref{sec53}.}
The marginal underperformance of \ours on MALOnt (0.5412 vs.\ 0.5647 for LLM4CTI) is consistent with our earlier observation that hybrid ontology discovery offers limited advantage on small schemas.
\textcolor{black}{The key findings from the ontology typing analysis are summarized as follows: (i) \ours generalizes to large hierarchical schemas where prompt-based schema inclusion baselines collapse, and (ii) predicate typing remains a harder task than entity typing for all four systems, motivating the ablation study on hybrid ontology discovery in Section~\ref{sec53}.}

\subsection{Ablation Study}
\label{sec53}

We isolate two design choices within hybrid ontology discovery: (i) the combination of embedding search and recursive schema navigation, and (ii) the embedding-similarity threshold $\tau$ that controls when the system switches from search to recursive traversal.
For both ablations, we use ground-truth entities and triplets as fixed inputs so that the reported scores reflect only the typing performance of each configuration, while all other parameters (ontology schema, candidate inventory, scoring rule) remain unchanged.

\subsubsection{Effect of Hybrid Ontology Discovery}
\label{sec531}

\textcolor{black}{We evaluate three configurations on both ontology typing tasks: Hybrid (\ours), Search-Only, and Recurse-Only. The Hybrid configuration combines embedding-based search with hierarchical recursive navigation, while the others rely solely on one of these strategies.}

\textcolor{black}{As shown in Table~\ref{tab:ablation_components}, the Hybrid configuration achieves the highest performance on both tasks, scoring 0.9364 on entity typing and 0.7843 on predicate typing, while the two single-strategy baselines exhibit asymmetric behavior.
On entity ontology typing, Search-Only (0.9184) approaches the Hybrid score, whereas Recurse-Only falls to 0.7914.
The pattern reverses on predicate ontology typing: Recurse-Only reaches 0.6416, while Search-Only collapses to 0.3142. Notably, the Hybrid configuration improves performance by 22.2\% (from 0.6416 to 0.7843) over Recurse-Only and more than doubles the score of Search-Only.}

\textcolor{black}{
The contrasting tendency of the two tasks reflects a structural difference between entity and predicate ontology typing.
Entity ontology typing maps a single named entity to a class, a problem that aligns well with semantic similarity over class names and descriptions, so embedding search alone suffices in most cases.
In contrast, predicate ontology typing requires reasoning over the subject-object relation, the property hierarchy, and directional constraints that distinguish subject from object, all of which are structurally represented in the schema and accessed through recursive navigation rather than lexical similarity.}
We further quantify this contrast in Appendix~\ref{appendix:transition}: instrumented measurements show that 91.0\% of entity-class queries are resolved by embedding search alone, while 72.1\% of ObjectProperty queries are routed to recursive navigation.
Ultimately, the hybrid configuration effectively combines these strengths: embedding-based search provides a fast lexical entry point for entities, and hierarchical recursive navigation handles the structural reasoning needed for predicates.

\begin{table}[t]
    \centering
    \renewcommand\arraystretch{1.0}
    \caption{Ablation Study on Hybrid Ontology Discovery Components}
    \begin{center}
    \begin{tabular}{lcc}
    \toprule
    \textbf{Configuration} & \textbf{Entity Ontology Typing} & \textbf{Predicate Ontology Typing} \\
    \midrule
    Search-Only       & 0.9184 & 0.3142 \\
    Recurse-Only      & 0.7914 & 0.6416 \\
    \textbf{Hybrid (ours)} & \textbf{0.9364} & \textbf{0.7843} \\
    \bottomrule
    \end{tabular}
    \label{tab:ablation_components}
    \end{center}
\end{table}

\subsubsection{Effect of Embedding Threshold}
\label{sec532}

To evaluate the effect of the embedding thresholds, we analyze $\tau_{\text{entity}}$ for class search in entity ontology typing and $\tau_{\text{predicate}}$ for property search in predicate ontology typing.
Each threshold controls when the system accepts an embedding search result rather than triggering a fallback to recursive navigation.
The system accepts a candidate only if its similarity score reaches or exceeds the corresponding threshold.
We sweep each threshold over $\{0.20, 0.25, 0.30, \ldots, 0.70\}$ while keeping all other parameters fixed.
We then measure the performance of the corresponding typing task.

\begin{figure}[t]
    \centering
    \subfigure[]{
        \includegraphics[width=0.85\columnwidth]{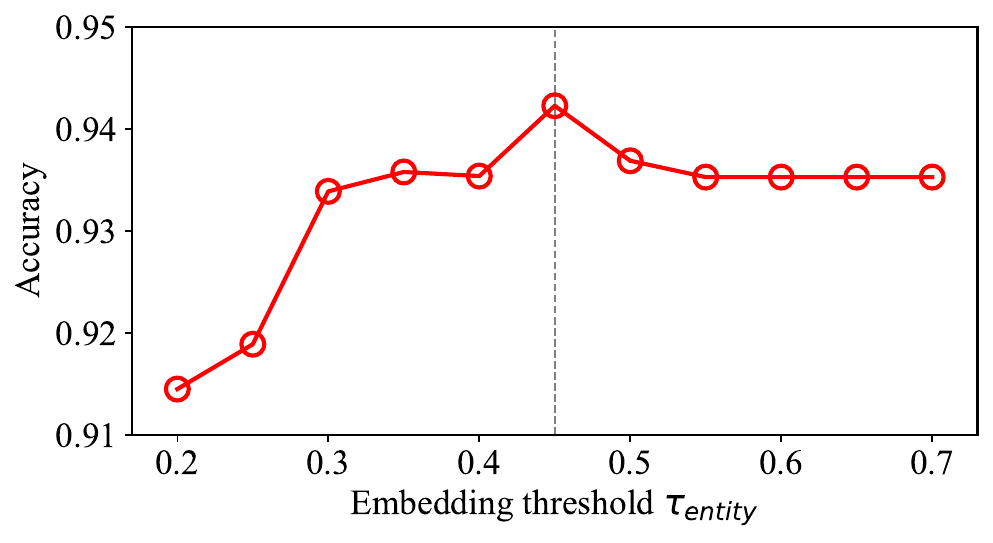}
        \label{fig:threshold_entity}
    }
    \subfigure[]{
        \includegraphics[width=0.85\columnwidth]{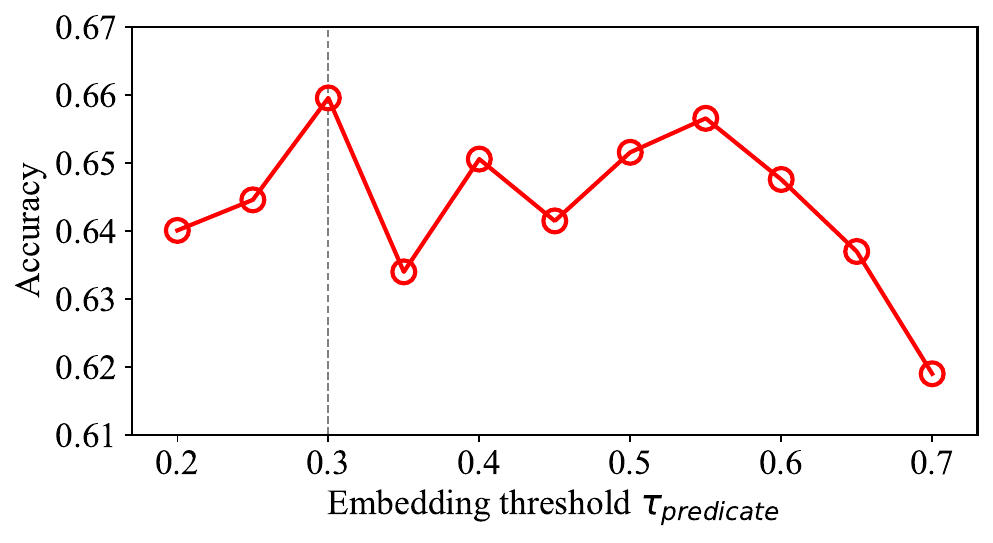}
        \label{fig:threshold_predicate}
    }
    \caption{
    Effect of the embedding thresholds on ontology typing performance: (a) entity ontology typing and (b) predicate ontology typing.
    }
    \label{fig:threshold}
\end{figure}

\textcolor{black}{As shown in Fig.~\ref{fig:threshold}, the two tasks exhibit distinct optimal thresholds and sensitivity patterns.}
For entity ontology typing, performance peaks at $\tau_{\text{entity}} = 0.45$ (0.9423) and remains within 0.7\% of this peak for $\tau_{\text{entity}} \in [0.30, 0.55]$.
This performance converges to a constant 0.9353 at $\tau_{\text{entity}} \geq 0.55$, where every query triggers recursive navigation.
For predicate ontology typing, performance peaks at $\tau_{\text{predicate}} = 0.30$ (0.6596) and is more sensitive to the threshold value.
This task shows moderate fluctuations in the middle range and experiences a sharper degradation by 6.2\% (from 0.6596 to 0.6190) at $\tau_{\text{predicate}} = 0.70$.

\textcolor{black}{This difference in optimal thresholds stems from the distinct embedding similarity distributions of the two search targets.}
As detailed in Appendix~\ref{appendix:transition}, entity-class queries demonstrate a high median similarity of 0.837, whereas ObjectProperty queries demonstrate a much lower median of 0.244.
Class names and descriptions provide rich lexical signals, enabling the system to tolerate a stricter threshold for entity ontology typing.
In contrast, property predicates offer weaker lexical signals, requiring a more permissive threshold to prevent the system from indiscriminately routing every query into recursive traversal.

\textcolor{black}{Consequently, we adopt $\tau_{\text{entity}} = 0.45$ and $\tau_{\text{predicate}} = 0.30$ as the default configurations for all subsequent experiments.
Note that optimal values may vary depending on the target ontology and downstream task.
The key findings from this ablation study are summarized as follows: (i) embedding-based search and hierarchical recursive navigation are fundamentally complementary, where embedding search drives entity ontology typing and recursive navigation drives predicate ontology typing, and (ii) the asymmetric lexical richness between classes and properties dictates distinct similarity thresholds for optimal performance.}

\subsection{Schema Compliance}
\label{sec54}

To investigate whether the constructed knowledge graph follows the formal structural rules of the target ontology, we measure the non-compliance rate over the full set of extracted entities and triplets on UCO.
This measurement is essential because downstream RDF/OWL reasoning systems reject malformed inputs at load time.
An item fails if it violates at least one of two checks: (i) URI and namespace conformance, and (ii) SHACL constraints declared by the schema.
These constraints cover required-attribute counts, datatype rules, malformed URIs, and missing required properties.

As shown in Table~\ref{tab:schema_compliance}, \ours achieves the lowest non-compliance rate of 5.2\%.
Specifically, \ours improves performance over the second-best baseline (CTINexus) by 88.2\% (from 43.9\% to 5.2\%).
CTINexus and LLM4CTI emit class-like labels that satisfy URI conformance for common entity types.
However, they skip schema-level checks, which causes missing required-attribute violations and datatype violations to dominate their failure counts.
TTPDrill produces rule-based templates with identifiers that diverge from the canonical UCO namespaces.
This divergence inflates URI conformance violations in addition to the required-attribute failures observed in the LLM-based baselines.

This performance gap arises from the closed-loop SHACL validation described in Section~\ref{sec44}.
The baseline systems~\cite{husari2017ttpdrill, cheng2025ctinexus, huangllm4cti} do not perform this validation.
\ours re-checks each predicted type against the required-attribute counts and datatype rules declared by the target schema before commitment.
The system re-invokes the discovery tool when it detects a violation.
We limit this self-correction to three retries to prevent unbounded execution.
The system records any remaining violation as a validation warning rather than silently dropping it.
Most violations in the LLM-based baselines are missing-attribute failures, such as a malware entity lacking the required hash algorithm attribute.
The closed-loop retry addresses these failures by routing a targeted property search back through the discovery tool.

\textcolor{black}{
The 5.2\% of non-compliant items in \ours represent violations that remain after the three-retry budget.
These items typically correspond to exceptional cases where no candidate property satisfies the missing required attribute.
\ours retains these items with their assigned class and a validation marker to preserve the extracted intelligence for downstream review.
A graph with 5.2\% marked items can be loaded into an RDF triplestore with minimal cleanup.
In contrast, the high violation rates of the baselines require pre-validation of every downstream query or manual cleanup of 40\% to 75\% of the items before reasoning is possible.}

\begin{table}[t]
    \centering
    \renewcommand\arraystretch{1.0}
    \caption{Schema Non-Compliance Rate on UCO Schema}
    \begin{center}
    \begin{tabular}{lccc}
    \toprule
    \textbf{System} & \textbf{Total Items} & \textbf{Violations} & \textbf{Non-compliance (\%)} \\
    \midrule
    TTPDrill~\cite{husari2017ttpdrill}  & 3,784  & 2,868 & 75.8 \\
    CTINexus~\cite{cheng2025ctinexus}   & 10,382 & 4,556 & 43.9 \\
    LLM4CTI~\cite{huangllm4cti}         & 3,881  & 1,950 & 50.2 \\

    \ours                               & 3,665  & 191   & \textbf{5.2} \\
    \bottomrule
    \end{tabular}
    \label{tab:schema_compliance}
    \end{center}
\end{table}

\subsection{Local LLM Feasibility}
\label{sec55}

A primary design goal of \ours is the integration of external CTI with sensitive internal incident data without exposing the data to enterprise LLM APIs.
To demonstrate this capability, we evaluate whether \ours preserves competitive typing performance with locally hosted open-source backbones.
We compare four LLM backbones under two conditions: with \ours and without \ours (i.e., naive prompt-based ontology typing).
The local backbones are Qwen3.5-35B and Gemma-4-26B, served through vLLM on the local workstation.
The enterprise backbones are GPT-5.4-mini and Claude Haiku-4.5, accessed through their official APIs.
As in Section~\ref{sec53}, we use ground-truth entities and triplets as fixed inputs so that the measured performance reflects only the typing stage.

As shown in Table~\ref{tab:local_vs_enterprise}, \ours improves typing performance for every backbone.
Specifically, for Qwen3.5-35B, Gemma-4-26B, GPT-5.4-mini, and Claude Haiku-4.5, \ours improves entity ontology typing by 118.6\% (from 0.4284 to 0.9364), 135.5\% (from 0.3905 to 0.9198), 107.3\% (from 0.4540 to 0.9410), and 74.4\% (from 0.5412 to 0.9439), respectively.
Predicate ontology typing exhibits the same pattern for all four backbones. 
The largest absolute improvements appear on local backbones, where the naive prompt-based scores are the lowest (Table~\ref{tab:local_vs_enterprise}).
The strongest enterprise configuration (Claude Haiku-4.5 with \ours) reaches 0.9439 on entity ontology typing and 0.8021 on predicate ontology typing. 
Meanwhile, the strongest local configuration (Qwen3.5-35B with \ours) reaches 0.9364 and 0.7843. 
This local configuration retains 99.2\% ($=\frac{0.9364}{0.9439}$) and 97.8\% ($=\frac{0.7843}{0.8021}$) of the performance of the best enterprise model, respectively.

\begin{table}[t]
    \centering
    \renewcommand\arraystretch{1.0}
    \caption{
    Effect of \ours on Local and Enterprise LLM Backbones
    }
    \begin{center}
    \begin{tabular}{llcc}
    \toprule
    \textbf{Backbone} & \textbf{Discovery} & \textbf{Entity} & \textbf{Predicate} \\
    \midrule
    \multirow{2}{*}{Qwen3.5-35B (local)}
        & w/o \ours & 0.4284 & 0.1491 \\
        & w/  \ours & \textbf{0.9364} & \textbf{0.7843} \\
    \midrule
    \multirow{2}{*}{Gemma-4-26B (local)}
        & w/o \ours & 0.3905 & 0.2849 \\
        & w/  \ours & 0.9198 & 0.6657 \\
    \midrule
    \multirow{2}{*}{GPT-5.4-mini (enterprise)}
        & w/o \ours & 0.4540 & 0.1581 \\
        & w/  \ours & 0.9410 & 0.8012 \\
    \midrule
    \multirow{2}{*}{Claude Haiku-4.5 (enterprise)}
        & w/o \ours & 0.5412 & 0.4318 \\
        & w/  \ours & \textbf{0.9439} & \textbf{0.8021} \\
    \bottomrule
    \end{tabular}
    \label{tab:local_vs_enterprise}
    \end{center}
\end{table}

We observe that Qwen3.5-35B with \ours (entity typing of 0.9364 and predicate typing of 0.7843) outperforms Claude Haiku-4.5 without \ours (entity typing of 0.5412 and predicate typing of 0.4318) by $1.73\times$ and $1.82\times$, respectively.
This result indicates that the integration of \ours into a local backbone delivers a larger gain than upgrading the backbone from local to enterprise.
\textcolor{black}{The four backbones converge to a narrow performance band with \ours. 
This convergence occurs because the system offloads the heaviest part of the reasoning, such as the navigation of large class and property hierarchies under formal constraints, from the LLM to the schema itself. 
Consequently, the LLM is left with a small bounded selection problem.
This finding demonstrates that for schema-grounded reasoning tasks such as ontology typing, the design of the discovery pipeline matters more than the raw capability of the underlying LLM.}
The key findings of the feasibility analysis are summarized as follows: (i) \ours offloads structural reasoning from the LLM to the schema, which reduces the performance gap between local and enterprise backbones, and (ii) local deployment with \ours retains 99.2\% (entity) and 97.8\% (predicate) of the best enterprise LLM performance. 
This level of performance supports privacy-preserving CTI knowledge graph construction without a loss of typing fidelity.

\section{Related Work} \label{sec:related_work}

The automation of structured knowledge extraction from CTI reports has been an active research topic.
We organize existing studies into three categories: rule-based pipelines, classification-based approaches, and LLM-based methods.

\textit{Rule-based pipelines.}
Husari~\textit{et al.}~\cite{husari2017ttpdrill} proposed TTPDrill, which mapped threat actions to MITRE ATT\&CK patterns through natural language processing (NLP) over predefined templates.
Satvat~\textit{et al.}~\cite{satvat2021extractor} introduced EXTRACTOR, which built attack graphs through dependency parsing and heuristic rules.
While these systems demonstrated automated CTI extraction, they rely on hand-crafted rules that require manual updates when new attack patterns appear.

\textit{Classification-based approaches.}
Later studies adopted machine learning models for CTI extraction.
Alam~\textit{et al.}~\cite{alam2023looking} presented LADDER, a BERT-based entity classifier for attack pattern recognition.
Peng~\textit{et al.}~\cite{peng2025retrieval} proposed a retrieval-augmented named entity recognition (NER) pipeline with adaptive instructions.
Mouiche and Saad~\cite{mouiche2025entity} combined SecureBERT~\cite{aghaei2022securebert} with a BiLSTM joint extractor for entity and relation extraction.
Piplai~\textit{et al.}~\cite{piplai2020knowledge} applied NER and fused the results into the UCO knowledge graph.
These methods improved upon rule-based systems, but they remain bound to fixed type inventories defined during training.
This limitation reduces their applicability when ontology schemas change.

\textit{LLM-based methods.}
Recent work has explored LLM-based pipelines for CTI knowledge extraction.
Huang~\textit{et al.}~\cite{huang2024ctikg} proposed CTIKG, which used GPT-4 with in-context learning (ICL) to extract open-ended knowledge graphs from CTI reports while deliberately avoiding fixed ontology schemas.
Cheng~\textit{et al.}~\cite{cheng2025ctinexus} proposed CTINexus, which adopted a similar ICL approach but aligned extracted triplets to the MALOnt ontology through in-context schema inclusion.
However, this design does not scale to large ontologies such as UCO.
LLM4CTI~\cite{huangllm4cti} introduced a chunk-wise dual-context framework with GNN-based link prediction, but it relies on a fixed, custom-defined schema and does not support formal ontology alignment.
Other LLM-based systems for CTI knowledge graph construction (TRACE~\cite{xu2026trace}, CTI-Thinker~\cite{yang2026cti}, AttacKG+~\cite{zhang2025attackg+}, LLM-TIKG~\cite{hu2024llm}) and general-purpose sequence-to-sequence extractors (KnowGL~\cite{rossiello2023knowgl}, REBEL~\cite{cabot2021rebel}) remain bound to fixed schemas or require task-specific fine-tuning.
Furthermore, none of these systems addresses cybersecurity-specific ontology constraints.
None of the above systems validates type assignments against a formal schema.
This omission causes hallucinated mappings to enter the downstream knowledge graph without verification.

\textcolor{black}{In summary, prior systems either bind to a fixed ontology during pipeline design or rely on in-context schema inclusion. 
Furthermore, none of these systems verifies ontology typing against formal schema constraints. 
Therefore, \ours has addressed both limitations through schema-agnostic hybrid ontology discovery with SHACL-based validation.}



\section{Discussion}\label{sec:discussion}


In this section, we discuss the operational trade-offs of local deployment, the structural boundaries of static ontologies, and the rationale behind our ontology typing mechanism.

\vspace{0.5ex}
\noindent

\textbf{Practical deployment in security operations.}
The local-deployment configuration (Section~\ref{sec55}) directly addresses privacy regulations (e.g., GDPR, HIPAA) that restrict sensitive data sharing.
Our local backbones achieve nearly the same typing quality as enterprise LLMs, making this privacy-preserving setup a practical default rather than a performance compromise.
The primary trade-off is increased processing latency due to multiple LLM calls per entity. 
However, in threat intelligence operations where reports are processed asynchronously, this latency is a reasonable cost to maintain complete control over sensitive data.

\vspace{0.5ex}
\noindent

\textbf{Ontology schema augmentation.}
Schema-aligned extraction is inherently bounded by the target schema's expressiveness. 
For example, when encountering novel entities like newly registered CVEs, the schema often lacks a suitable class.
In these cases, \ours conservatively defaults to \textit{owl:Thing} rather than assigning an imprecise or fabricated type.
This fallback mechanism highlights a fundamental limitation of mapping dynamic text to a static ontology: a system cannot assign a concept that the schema does not define.
Future work will explore ontology schema augmentation methods, dynamically generating new classes for concepts that repeatedly trigger this fallback to relieve static schema constraints.

\vspace{0.5ex}
\noindent

\textbf{Confidence-based typing.}
\ours rejects mappings below specific confidence thresholds ($\tau_{\text{entity}}$ and $\tau_{\text{predicate}}$), defaulting unverified items to owl:Thing or generic properties. 
This prevents low-confidence guesses from corrupting the knowledge graph, which is essential to avoid misleading threat attribution in security operations.
While we establish default thresholds ($\tau_{\text{entity}} = 0.45$, $\tau_{\text{predicate}} = 0.30$) based on Section~\ref{sec53}, these are empirical baselines. 
Operators should recalibrate these values depending on the target ontology's complexity and the specific downstream analytical objective.

\section{Conclusion}
\label{sec:conclusion}

We have presented \ours, a schema-agnostic CTI knowledge graph construction framework.
\ours has addressed the limitations of existing LLM-based extraction pipelines, such as schema dependency, scalability issues on large ontologies, and the privacy risks of enterprise LLMs.
\ours introduces a \textit{hybrid ontology discovery} mechanism that combines embedding-based search with LLM-guided hierarchical recursive navigation.
Furthermore, a SHACL-based closed-loop validation enforces schema-compliant ontology typing, reducing hallucinated mappings in the output graph.
Experimental results on UCO, STIX, and MALOnt demonstrated that \ours improves UCO entity and predicate ontology typing by 62.5\% and 29.5\%, while decreasing the schema non-compliance rate from 43.9\% to 5.2\%.
When paired with a locally deployed open-source LLM, \ours maintains 99.2\% (entity) and 97.8\% (predicate) of the typing performance of the best enterprise LLM, supporting privacy-preserving CTI analysis without a loss of typing fidelity.
In future work, we will explore cross-ontology translation, where the schema-agnostic reasoning of \ours maps threat knowledge between disparate schemas to facilitate seamless inter-organization CTI sharing.

\section*{LLM Usage Statement}
LLMs were used for editorial purposes in this manuscript,
and all outputs were inspected by the authors to ensure
accuracy and originality.


\bibliographystyle{IEEEtran}
\bibliography{reference}



\appendices

\section{Details of Hybrid Ontology Discovery}
\label{appendix:transition}

The ablation study in Section~\ref{sec53} shows that entity ontology typing and predicate ontology typing draw on different components of hybrid ontology discovery.
We analyze this difference by recording, for each typing query, whether the final answer comes from the embedding-based search step or from the hierarchical recursive navigation step.
We also measure the top-1 search similarity to indicate how reliably embedding similarity resolves each phase.

\begin{table}[h]
    \centering
    \renewcommand\arraystretch{1.0}
    \caption{Typing Queries Resolved by Search vs. Recursive Navigation}
    \begin{center}
    \begin{tabular}{lcc}
    \toprule
    \textbf{Phase} & \textbf{Search} & \textbf{Recurse} \\
    \midrule
    Entity (class)    & 91.0\% & 9.0\%  \\
    DatatypeProperty  & 54.4\% & 45.6\% \\
    ObjectProperty    & 27.9\% & 72.1\% \\
    \midrule
    \textbf{Total}    & \textbf{62.3\%} & \textbf{37.7\%} \\
    \bottomrule
    \end{tabular}
    \label{tab:tool_resolution}
    \end{center}
\end{table}

\textbf{Resolution by phase.}
Table~\ref{tab:tool_resolution} shows that the two phases of hybrid ontology discovery are exercised at very different rates depending on the typing target.
The search-resolution rates for entity class, DatatypeProperty, and ObjectProperty queries are 91.0\%, 54.4\%, and 27.9\%, respectively,
while the remaining 9.0\%, 45.6\%, and 72.1\% are answered by recursive navigation.
Entity typing and ObjectProperty typing sit at opposite ends of this routing: the search-resolution rate drops by 69.3\% (from 91.0\% to 27.9\%) between the two phases.
DatatypeProperty typing sits between the two extremes at 54.4\% search and 45.6\% recurse, because literal-valued properties carry stronger lexical signals than relations but weaker structural cues than entity classes.
Aggregated over the three phases, 62.3\% of typing queries are answered by search and 37.7\% by recursive navigation, confirming that neither component dominates the workload.

\begin{table}[h]
    \centering
    \renewcommand\arraystretch{1.0}
    \caption{Top-1 Embedding Similarity and Threshold Clearance Rates}
    \begin{center}
    \begin{tabular}{lccc}
    \toprule
    \textbf{Phase} & \textbf{Median} & \textbf{Mean} & \textbf{Above $\tau$} \\
    \midrule
    Entity (class)    & 0.837 & 0.856 & 92.9\% \\
    DatatypeProperty  & 0.507 & 0.512 & 60.0\% \\
    ObjectProperty    & 0.244 & 0.300 & 13.2\% \\
    \midrule
    \textbf{All}      & \textbf{0.624} & \textbf{0.702} & \textbf{57.9\%} \\
    \bottomrule
    \end{tabular}
    \label{tab:similarity_stats}
    \end{center}
\end{table}

\textbf{Embedding similarity distribution.}
As shown in Table~\ref{tab:similarity_stats}, the top-1 similarity statistics align with the routing pattern in Table~\ref{tab:tool_resolution}.
The median top-1 similarities for entity-class, DatatypeProperty, and ObjectProperty queries are 0.837, 0.507, and 0.244, respectively, and the means follow the same ordering at 0.856, 0.512, and 0.300.
The threshold-clearance rates differ even more sharply: 92.9\% of entity-class queries exceed $\tau_{\text{entity}} = 0.45$, whereas only 13.2\% of ObjectProperty queries exceed $\tau_{\text{predicate}} = 0.30$.
ObjectProperty typing depends on the domain-range pair of subject and object entities rather than on the surface verb, so embedding similarity over surface forms is a weak signal, which is consistent with the drop in median top-1 similarity from 0.837 for entity-class queries to 0.244 for ObjectProperty queries.
DatatypeProperty queries clear the predicate threshold at 60.0\%, sitting between the two extremes and matching the intermediate 54.4\% search-resolution rate in Table~\ref{tab:tool_resolution}.

\textbf{Explanation of the ablation study.}
Together, Tables~\ref{tab:tool_resolution} and~\ref{tab:similarity_stats} explain why the ablation study in Section~\ref{sec53} affects entity typing and predicate typing so differently.
Search-Only performs well on entity ontology typing because 92.9\% of entity-class queries already clear $\tau_{\text{entity}}$ by embedding similarity, but it loses most of predicate ontology typing because only 13.2\% of ObjectProperty queries clear $\tau_{\text{predicate}}$ and the remaining 86.8\% have no recursive navigation route available.
Recurse-Only partially recovers the ObjectProperty case because recursive navigation can travel the domain-range structure, but it underperforms on entity ontology typing because it discards the median similarity of 0.837 lexical signal in favor of a longer LLM-driven navigation.
The full hybrid configuration (\ours) combines the two complementary signals: search resolves the lexically dominated portion (62.3\% of all queries), and recursive navigation handles the schema-driven remainder (37.7\%).

\end{document}